\documentclass[usenatbib,times]{mn2e}
\usepackage{graphicx}
\usepackage{epsf}

\usepackage{url}

\def\le{{L_{\rm Edd}}}

\def\fermi{{\it Fermi}}
\def\gbm{{{\it Fermi}-GBM}}
\def\lat{{{\it Fermi}-LAT}}

\def\H0{{\rm ~km~s^{-1}~Mpc^{-1}}}

\begin{document}

\title[A high-energy peak in GRB 080825C]{Signs of magnetic acceleration and multi-zone emission in GRB 080825C}

\author[Moretti \& Axelsson]{Elena Moretti,$^{1,2}$\thanks{email: moretti@mpp.mpg.de, emoretti@kth.se}
and Magnus Axelsson$^{2,3}$\\
$^{1}$Max-Planck-Institut f\"ur Physik, F\"ohringer ring 6, D-80805 M\"unchen, Germany\\
$^{2}$Oskar Klein Center for CosmoParticle Physics, KTH Royal Institute of Technology, SE-106 91 Stockholm, Sweden\\
$^{3}$Department of Physics, Tokyo Metropolitan University, Minami-osawa 1-1, Hachioji, Tokyo 192-0397, Japan\\
}

\date{Accepted --. Received --; in original form --}

\pagerange{\pageref{firstpage}--\pageref{lastpage}} \pubyear{2002}

\maketitle

\begin{abstract}
One of the major results from the study of gamma-ray bursts with the {\it Fermi} Gamma-ray Space Telescope has been the
confirmation that several emission components can be present in the energy spectrum. Here we reanalyse the spectrum of GRB 080825C 
using data from the {\it Fermi} LAT and GBM instruments. Although fairly weak, it is the first gamma-ray burst detected by the {\lat}. 
We improve on the original analysis by using the 
LAT Low Energy (LLE) events covering the 30--100 MeV band. We find evidence 
of an additional component above the main emission peak (modelled using a Band function) with a significance of $3.5\sigma$ in 
2 out of the 4 time bins.
The component is well fitted by a Planck function, but shows unusual behaviour: the peak energy increases in the prompt emission 
phase, reaching energies of several MeV. This is the first time such a trend has been seen, and implies that the origin of this component 
is different from those previously detected. We suggest that the two spectral components likely arise in different regions of the outflow, and 
that strong constraints can be achieved by assuming one of them originates from the photosphere. The most promising model appears 
to be that the high-energy peak is the result of photospheric emission in a Poynting flux dominated outflow where the magnetisation
increases with time.

\end{abstract}

\section{Introduction}

The launch of the {\fermi} gamma-ray satellite has led to much progress in the study of gamma-ray bursts (GRBs).
Spectral features hinted at in the {\it CGRO} BATSE and EGRET data have been fully confirmed and their properties further studied:
for example the presence of an extra component modeled as a power law rising at high energies and lasting for longer than the
bulk of the keV--MeV emission \citep{Gonzalez03,catalog}; or a low-energy (tens of keV) thermal component modelled with a 
Planck function \citep{ryde04,axe12}.
Moreover, new spectral features have been discovered thanks to the unprecedented energy coverage for transient events
(8 keV to more than 300 GeV) of {\fermi} and to its timing capabilities. For example in the case of GRB 090926A \citep{ackermann11} 
the high-energy power-law component shows a significant cut-off at about 1.5 GeV. In the case of GRB 100724B a cut-off of the Band 
function \citep{band} high-energy slope has been observed at about 40 MeV \citep{catalog}. 

While extra components can be detected in bright bursts, as the photon statistics decrease so does the statistical significance of 
such deviations from the Band function. The results from the LAT \citep[Large Area Telescope,][]{Atwood09} GRB catalog 
\citep{catalog} demonstrate that the brighter 
bursts indeed have more complex spectra with respect to the simple Band function. In dimmer bursts, the presence of these 
extra components cannot be excluded, as they might not be detectable due to inadequate photon statistics. The search for extra
components in the dimmer bursts will therefore result in less significant detections, making the identification of such spectral features
more difficult. It is not clear whether intrinsically weak GRBs are the result of different conditions in the surrounding medium, lower 
energy input from the central engine, or whether they reflect differences in the relativistic outflow itself, making such searches 
important. For the LAT GRB catalog, a lower limit of $4.2\sigma$ (corresponding to a chance 
probability $< 10^{-5}$) was set in order to claim a significant deviation from a pure Band function.

In order to search for features with lower significance we initiated a systematic re-examination of the LAT-detected GRBs, 
to look for features with lower significance that may have been missed in previous searches due to the lack of photon statistics. 
In this paper we present the results from the first such reanalysis, of GRB 080825C. It is the first LAT-detected dim burst and was chosen 
merely on account of its detection date; no other selection criteria were applied. It is a faint burst which has previously 
been presented in \citet{abdo09}. There, and in the LAT GRB catalog, the spectrum was found to be adequately described by a Band 
function. We begin by describing the observations and our data analysis in Sect.~2. Thereafter we present our results in Sect.~3. 
Finally, we discuss our findings.

\section{Observations and data analysis}

GRB 080825C is the first GRB detected by the LAT. It was triggered at 4:13:48 UTC on 2008 August 25
from the {\gbm} \citep[Gamma-ray Burst Monitor,][]{gbmpaper} flight software. On the basis of a dedicated off-line analysis, the GBM 
team localized this burst at 
RA = $232.2^{\circ}$, Dec = $-4.9^{\circ}$ (J2000), with a statistical uncertainty of $1.5^{\circ}$ and a standard systematic uncertainty of 
2$^{\circ}$--3$^{\circ}$. The $T_{90}$ duration (the time during which 90\% of the flux is received) estimated from the GBM data 
(8--1000 keV) is about 27 s, although after 25 s the emission detected from the BGO detectors is extremely weak. In the NaI detectors, 
as well as in the LAT, emission is seen up to 35 s \citep{abdo09}. 

In our analysis we use the data from both the GBM and LAT instruments. The detectors used, the definition of the time bins 
and the energy intervals are the same as in the previous studies \citep{abdo09,catalog}. However, in our spectral analysis we 
also include a new type of data-set, LAT Low Energy Pass 7 (LLE) events \citep{lleref,Aje14}. The LLE data-set is obtained by applying different 
(somewhat looser) selection criteria on the reconstruction quality of the LAT events. It is thereby possible to lower the energy range 
of the spectral analysis in the case of transient events like GRBs, where the signal-to-noise ratio is favourable. 
The standard LAT data are selected with Pass 7 reconstruction and classification \citep{p7ref}. We use the TRANSIENT class of Pass 7 reprocessed 
data\footnote{\url{http://fermi.gsfc.nasa.gov/ssc/data/analysis/documentation/Pass7REP_usage.html}.}.
The LAT data were selected within a region of interest of 12 degrees 
around the location of the burst together with a zenith angle cut of 100 degrees and a low energy cut of 100 MeV to not overlap 
with the LLE data.
The latest GBM data products (time-tagged event, TTE) and response files are used for the spectral analysis in the keV region. Following 
\citet{abdo09} we use NaI detectors 9 and 10 and both BGO detectors. The background estimation for the GBM and the LLE data-sets were made
following the procedure described in \citet{catalog}, and for the standard LAT data the Background Estimator \citep{BKGE} was used. The 
spectral analysis and the simulations to calculate the significance were performed with the {\sc xspec} package, v12.8.2.

Again following \citet{abdo09}, the GRB light curve was divided into 5 time bins from 0.0 s to 35.45 s. For each time bin a dedicated spectral 
analysis was performed to compare a fit using the Band function only to one with a Band function plus extra ``components'': cut off power law, 
Planck function or an extra Band function. As test statistic we used {\tt pgstat}\footnote{PG-statistics are used in the case
of a Poissonian distributed signal and a Gaussian distribution of the background (full details can be found in the {\sc xspec} manual).}.  

Due to the limitations arising from poor count statistics, we accept a lower significance than \citet{catalog} when 
considering the possibility of extra spectral components, and investigate deviations $>3\sigma$. The statistical results are reported 
together with a Band-only fit for comparison. The parameters of this model agree with the previously published ones in the first two time bins, 
while the results in the subsequent time bins do not match those of \citet{abdo09}. The discrepancy is greatest for the $\beta$ parameter,
corresponding to the high-energy index of the Band function. This change is a result of adding the LLE data, which help to better
constrain this part of the spectrum.

\section{Results}

In the first and the third time bins there is no emission in the LAT above 100 MeV; however, we do find emission in the LLE range. In the 
last bin there is only very weak emission in the BGO detectors, but significant emission above 100 MeV. The time-resolved spectral analysis 
of \citep{abdo09} identifies the Band function as the best fit in all the time bins but the last, where a power law gave the best fit. 

As in \citet{abdo09} we find that a single power law is statistically preferred over the more complex Band function in the last time bin.
However, in our analysis we also find signs of a deviation from a Band function in the first and fourth time bins. In particular, 
there appears to be an
additional feature in the spectrum at high energies (above 800 keV). We therefore add a Planck function in our model to capture this
peak. This function is physically motivated as it expresses a thermal emission spectrum, and has been used to model photospheric emission 
at low energy \citep[for example][]{axe12}. However, in this first step our goal is to capture the extra peak without making assumptions 
about its origin; possible interpretations will be discussed in Sect.~\ref{disc}. A Planck function is a simple function with only two degrees 
of freedom, which allows us to model the spectral feature with the fewest free parameters. 

\begin{figure*}
\includegraphics[width=8cm]{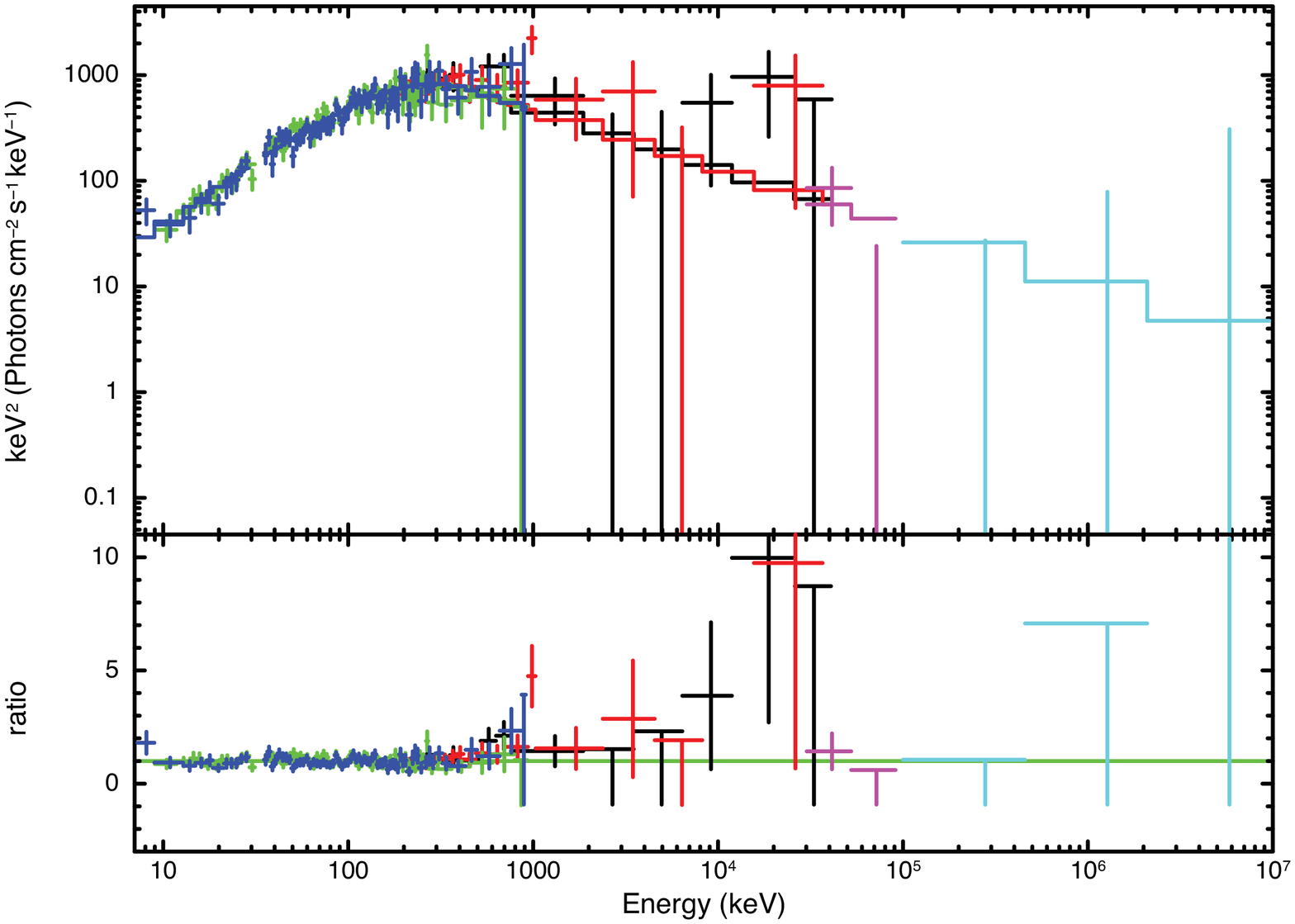}
\hspace{1cm}
\includegraphics[width=8cm]{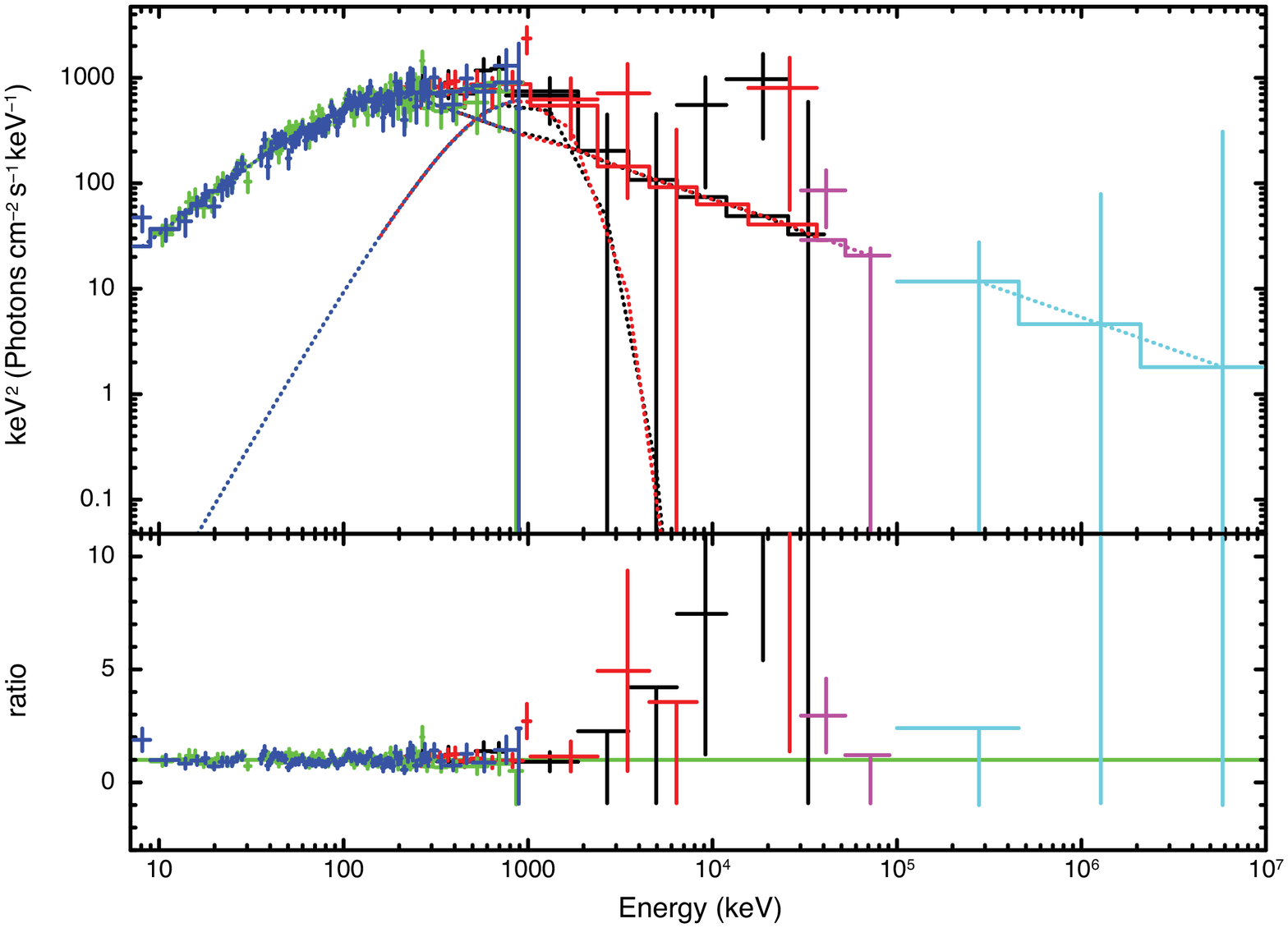}
\caption{Results of the spectral fit of the first time bin, 0--2.69\,s, using a single Band function (left) and a Band function with an 
additional Planck function (right). Panels show the deconvolved model spectrum (top), the model count spectrum including data points 
(middle) and ratio of the residuals to the model (bottom). Data points are colour-coded according to detector: NaI (green and blue), BGO (red and black), 
LLE (magenta) and LAT (cyan). In the ratio panel, upper limits have been extended down to -1 for clarity.}
\label{bin1}
\end{figure*}

\begin{figure*}
\includegraphics[width=8cm]{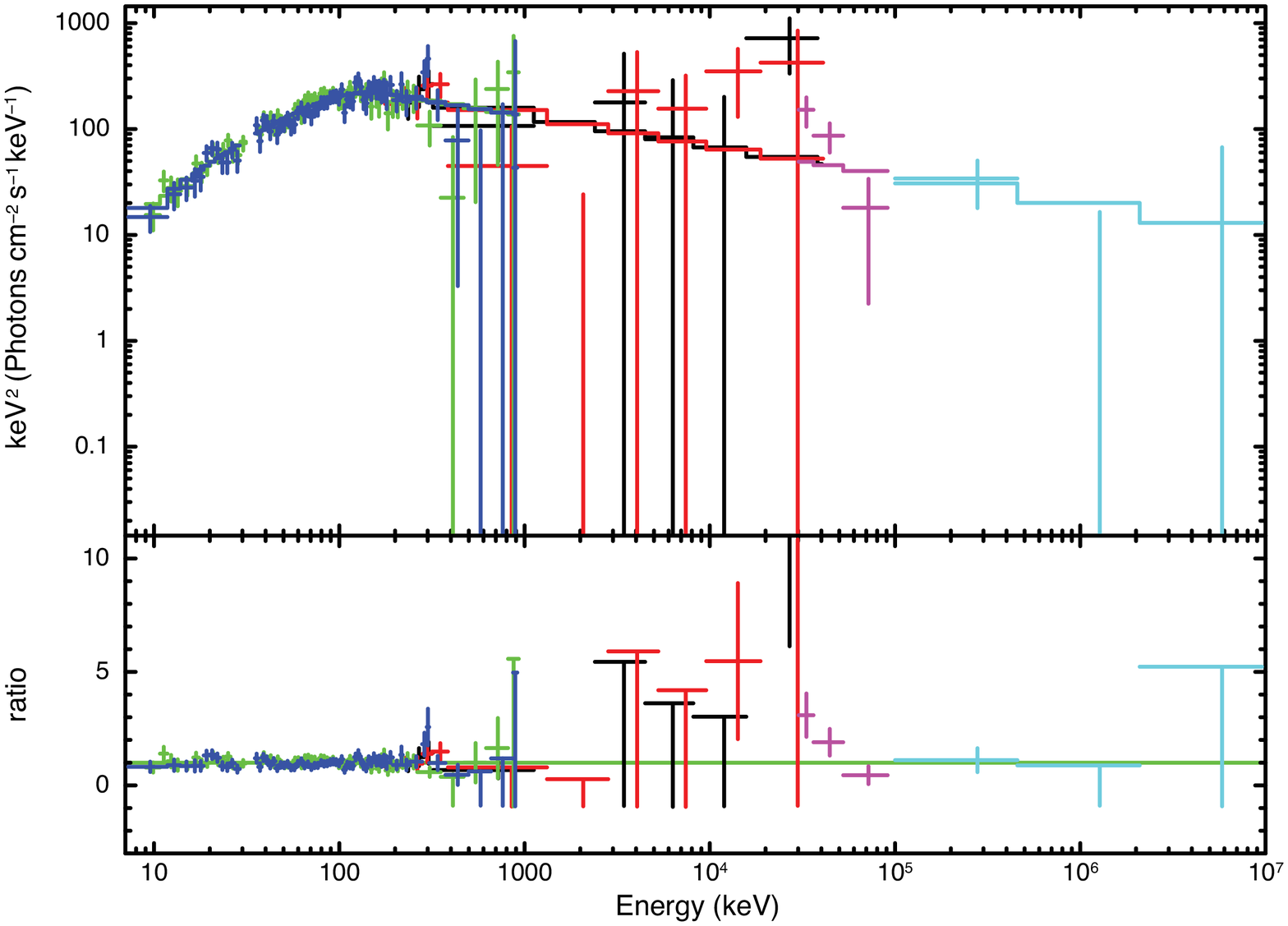}
\hspace{1cm}
\includegraphics[width=8cm]{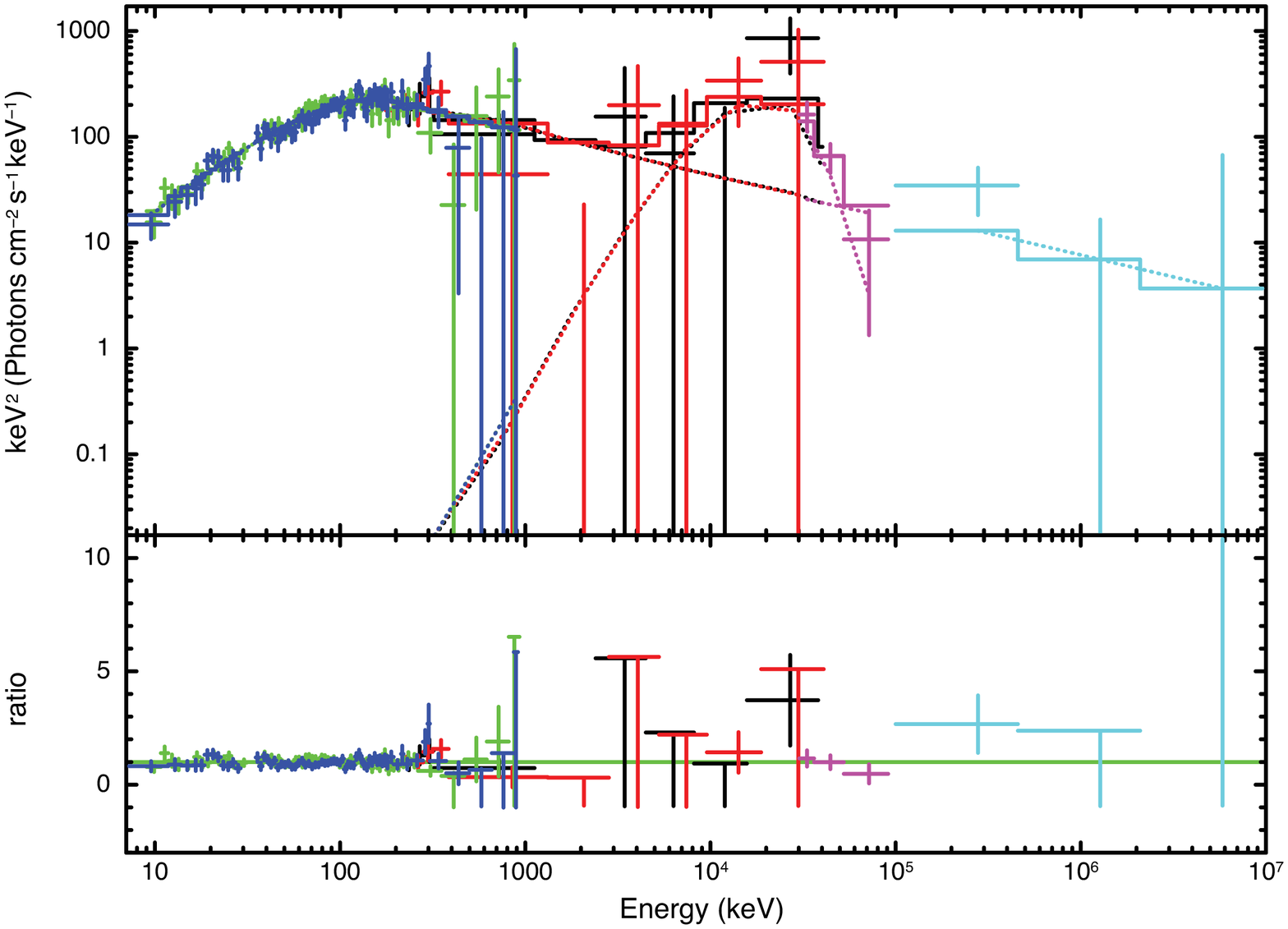}
\caption{Results of the spectral fit of the fourth time bin, 12.93--25.22\,s, using a single Band function (left) and a Band function with an 
additional Planck function (right).
Panels show the deconvolved model spectrum (top), the model count spectrum including data points (middle) and ratio of the residuals to the model  (bottom). Data 
points are colour-coded according to detector: NaI (green and blue), BGO (red and black), LLE (magenta) and LAT (cyan). In the ratio panel, upper limits have been 
extended down to -1 for clarity.}
\label{bin4}
\end{figure*}

Determining the presence of an extra component by comparing values of {\tt pgstat} (or other test statistics) is not straight-forward 
\citep[see, e.g., the discussion in][]{pro02}, so to derive an estimate of significance from the $\Delta${\tt pgstat} values we use Monte Carlo 
simulations. For each Band function fit, $10^5$ spectra were generated and fit with a single Band function as well as with a Band function 
plus a Planck function. The resulting distribution of $\Delta${\tt pgstat} values were then used to derive the chance probability to see the 
improvement obtained in the actual fit to the data. We find that the significance of the extra component is $3.5\sigma$ in the first and fourth 
time intervals. These fits are shown in  Figs.~\ref{bin1} and \ref{bin4}.

The full results of the time-resolved spectral fits can be found in Table~\ref{fitresults}. The fact that the component is seen in both early and late stages of the burst
motivates us to include an additional Planck function in all time bins. With respect to the first and the fourth time bins, in the second and third time 
bins the significance of the extra component drops below $1\sigma$. This is likely due to the energy region the component passes through at these times. In the second and third time 
bins the second spectral peak is found in the region of 2--5 MeV, where only the BGO detectors are sensitive. Unfortunately this is also the energy range where the GBM 
effective area drops significantly. Therefore the lower sensitivity in this region hampers the search for significant extra components. In the 
fourth time bin the second peak appears between the BGO and the LLE sensitivity regions, around 15 MeV. The use of the LLE data is thus important to 
establish the presence of the component at later times. We note that although the significance of the extra component is not the same in all 
time bins, its parameters are constrained throughout (Table~\ref{fitresults}). 
In addition, the smooth evolution of the extra component decreases the risk that it is a spurious detection.

\begin{table*}
\begin{center}
\caption{Results of the time-resolved fits where the first four bins are modelled with a Band plus black body functions while the fifth time bin is 
modelled with a simple power-law function. In the table $\alpha$ and $\beta$ are the low and high-energy indexes, $E_{p}$ is the peak energy of the 
Band function and $N_{Band}$ its normalisation. $kT$ and $N_{BB}$ refer to the black body function and are, respectively, the temperature and the 
normalisation, while $N_{PL}$ is the normalisation of the power-law function. \label{fitresults}}
\begin{tabular}{cccccccccc}
\hline \hline
Time (s) & $\alpha$ & $\beta$ & $E_{p}^{a}$ & $N_{Band}^{b}$ & $kT^{a}$ & $N_{BB}^{b}$ & $N_{PL}^{b}$ & $PG_{stat}$/DoF & $\Delta PG_{stat}^{c}$ \\
\hline
0.0 -- 2.69 &$-0.56^{+0.08}_{-0.07}$ & $-2.6^{+0.1}_{-0.2}$ &  $203^{+31}_{-28}$  &  $0.10^{+0.01}_{-0.01}$ &  $219^{+62}_{-36}$ &  $15^{+3}_{-3}$ & - & 523/493 & 16\\
2.69 -- 4.74 &$-0.46^{+0.06}_{-0.07}$ & $-2.43^{+0.06}_{-0.07}$ &  $208^{+22}_{-19}$  &  $0.14^{+0.01}_{-0.01}$ &  $632^{+208}_{-221}$ &  $11^{+8}_{-7}$ & - & 496/493 & 3\\
4.74 -- 12.93 &$-0.74^{+0.05}_{-0.05}$ & $-2.46^{+0.07}_{-0.08}$ &  $183^{+19}_{-17}$  &  $0.046^{+0.004}_{-0.003}$ &  $1191^{+497}_{-332}$ &  $7^{+4}_{-4}$ & - & 505/493 &  4\\
12.93 -- 25.22 &$-0.64^{+0.05}_{-0.05}$ & $-2.42^{+0.06}_{-0.06}$ &  $151^{+13}_{-13}$  &  $0.050^{+0.004}_{-0.004}$ &  $4613^{+920}_{-508}$ &  $6.0^{+2.0}_{-1.5}$& - & 541/493 & 15\\
25.22 -- 35.46 &$-1.94^{+0.03}_{-0.04}$ & - &  -  & - &  - &  - & $8.3^{+1.0}_{-1.4}$ &464/499 & -8\\
\hline
\end{tabular}
\end{center}

\flushleft
{$^a$}{keV}\\
{$^b$}{ph cm$^{-2}$ s$^{-1}$ keV$^{-1}$}\\
{$^c$}{To convert a given change in $PG_{stat}$ to a significance value, simulations must be performed as described in Sect.~3.}

\end{table*}

Since we have lowered our threshold for accepting additional components, the effects of systematic errors are a concern. 
We therefore investigated their impact on the spectral fits. To test the dependence of the result on the response 
function of the GBM instruments we used response matrices generated for locations three degrees away from the GRB position.
As the component lies  between different detectors in the first (NaI--BGO) and fourth (BGO--LLE) time bins we also investigated the impact 
of the relative difference in the detector effective areas. Independently increasing or decreasing the effective area of the 
GBM-NaI, GBM-BGO and LLE data sets by 15\% did not give any strong effect. In all the above tests, the fitted parameter values remained in 
agreement with 
the original ones, and the significance of the extra component did not change more than marginally (the lowest significance measured 
was $3.2\sigma$). Discussion with the {\it Fermi}-GBM team indicated that other effects on the energy dispersion and calibration were likely 
below the statistical errors. The energy calibration is checked by the GBM team by the means of emission lines of the background and 
of solar flares \citep{gbmpaper}. The energy dispersion was studied in the on-ground calibrations \citep{Biss09}.
Additionally, we see the feature in several different detectors and in different time bins. We therefore conclude that it is unlikely to 
be caused by systematic effects.

Because of the unusually high energy of the second peak, and the fact that the Planck function has hitherto only been seen below the Band 
function peak, we 
investigate the possibility that the two components have switched position in our fits, with the Planck function capturing the peak normally described by
a Band function.
However, we find that it is not possible to change places of the two components - the peak at lower energy is too wide to be well 
described by a Planck function. On the other hand we note that the second peak is compatible with a wider fitting function. It is
for example possible to fit the spectra with two Band functions, indicating that the width of the second peak is not necessarily as 
narrow as a Planck function. Unfortunately, the limited statistics mean that we are unable to fully constrain 
the parameters of more complex models, such as two Band functions. 
Although we do not report these fits it is important to note that the high-energy peak can be described by a wider function than the Planck one.

\begin{figure}
\includegraphics[width=8cm]{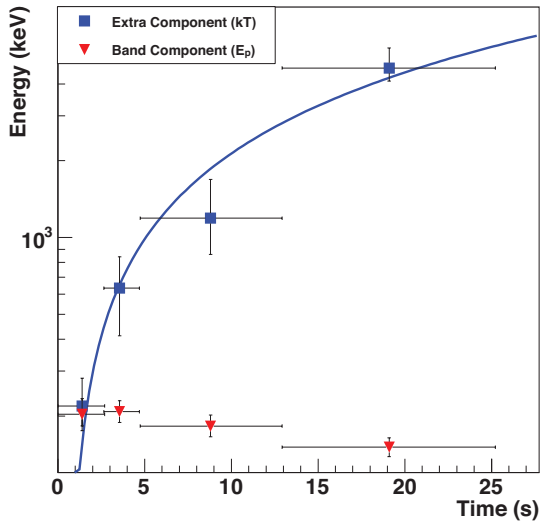}
\caption{Temporal evolution of the Band and Planck peak energies. The blue line illustrates a linear relationship (not fit to the data).}
\label{evolution}
\end{figure}

Looking at the temporal evolution throughout the burst, it is clear that the Band component and the extra peak show different behaviour.
The evolution of the two peaks is shown in Fig.~\ref{evolution}. While the low- and high-energy indices of the Band function remain stable in the first four 
time bins ($\alpha \sim -0.6$ and $\beta \sim -2.5$), the $E_p$ shows a sign of decreasing towards the end of the burst (matching the standard behaviour). 
The parameters of the extra component evolve throughout the emission phase: the energy of the second component increases sharply throughout the burst, 
reaching a value of $kT>4$\,MeV in the fourth time bin, while the normalisation decreases.

\section{Discussion}
\label{disc}

As presented in the previous section, we find evidence for an extra peak in the spectrum of GRB 080825C. While the significance is $3.5\sigma$ 
only in the first and fourth time bins, the fact that the recovered peak energy follows a linear relation (blue line in Fig.~\ref{evolution}) for all bins suggests that 
this spectral component is actually present during the full burst emission. A component showing such behaviour has never been observed before: 
the spectral characteristics and its time evolution are dissimilar to the standard hard-to-soft peak evolution \citep{kan06,rp09}. As this is the first
GRB in our sample to be investigated, we cannot rule out the possibility of other bursts having 
similar components. The population studies made so far \citep[e.g.,][]{catalog} have not included any doubly-peaked model (e.g., Band+blackbody). Combined 
with the low number of bursts included (35), this means that it is difficult to draw any firm conclusions regardless of the intensity of the GRB.

Previous studies have found extra spectral components in spectra of bright GRBs, such as GRB 110721A \citep{axe12}. However, while these components
are also typically modelled with Planck functions, they appear at much lower energies ($\leq200$ keV). They are ascribed to photospheric emission, and
the temperature typically decreases with time as a broken power law. The peak of this component increases with time, in contrast to the standard 
hard-to-soft peak evolution \citep{kan06,rp09}.

The high-energy peak in GRB 080825C is also different from the typical Band function seen in other GRBs. For instance, the peak energy generally
decreases with time, such as in GRB 110721A \citep{axe12} or GRB 130427A \citep{pre13}. The different temporal behaviour of the highest-energy 
peak in GRB 080825C may therefore suggest a different physical origin, for example Inverse-Compton (IC) cooling as suggested 
in \citet{vurm14}. In addition to the deviating temporal behaviour, the second peak also reaches very high energies. Indeed, the energy of this peak in the 
fourth time bin is $\sim 15$ MeV, which is very close to the value GRB 110721 had in its {\em first} time bin ($\sim15$ MeV). Such high energies are similar to the 
modeled Inverse Compton component in the slow heating scenario presented in \citet{peer06}. 

If there are two components in the spectrum of GRB 080525C, it is possible to attribute them to two separate emission components. The question
then arises if the emission comes from the same region, or two separate regions. 
In the case the emission comes from the photospheric radius we can distinguish two different scenarios. In the scenario of no dissipation below the 
photospheric radius the black body component of the last bin has too high temperature (in relation to its luminosity) to be the thermal 
emission \citep[for example][]{hascoet}. Thus the Band component must be tied to 
thermal emission, leaving the black body component to be explained. 
On the other hand, if there is dissipation below the photosphere, in particular from multiple radiation-mediated shocks, a double-peaked spectrum could be 
produced similar to the one \citet{Levinson14} find. However, we note that the observed temperature evolution of the high-energy peak does not match 
the one predicted by \citet{Levinson14} of a strong shock followed by weaker ones.
We therefore conclude that the mismatch disfavours this interpretation of the data together with the 
possibility the emission solely comes from the photospheric radius. 

In the context of a two-zone scenario it is not clear which component is related with the photosphere. If the second peak is chosen, the high temperature 
points to the presence of magnetic dissipation in the outflow, because in the case of pure thermal emission, high temperature is correlated with 
high radiative efficiency \citep{hascoet}. In the case of a Poynting flux dominated outflow, the high-energy peak can thus be interpreted as photospheric emission in the 
presence of magnetic dissipation below the photosphere \citep[as in][]{bp15}. 
This would lead to a high temperature of the peak and give low efficiency, providing only a few percent of the burst total emission
which matches the observation. 
Strongly magnetised outflows lead to a suppression of the photon production, which leads to the thermal peak reaching very high energies, 
above 8 MeV. Depending on the details of the energy deposition, the spectrum can attain shapes that are very different from, and wider 
than, a Wien or a Planck spectrum \citep[for example,][use this scenario to explain the main Band-function component
in GRB 110721A]{bp15}.
We note that the blackbody temperature at the beginning of the emission is lower than in the last bins. In the early
phases the magnetic dissipation would therefore be low, and the temperature more similar to a pure photospheric outflow without dissipation. 
The emission of the outflow can vary with time according to the conditions of the environment, the central engine activity and the dynamics of the 
emitting shells. Subsequent shells can be emitted under different conditions and therefore carry different properties like the amount of magnetisation
and the bulk Lorentz factor \citep[see][]{hascoet}. Magnetically dominated outflows are not expected to have shocks, but the Band component could be 
explained through synchrotron emission of electrons accelerated via reconnection \citep[such as the ICMART scenario of][]{zhang}. 

It is also possible to tie the Band component to the photosphere. The high efficiency then indicates that transparency occurs in the
acceleration phase, possibly in the transition between coasting and acceleration. The challenge is however to explain the blackbody component.
In the scenario of inverse Compton emission suggested by \citet{bel14}, a delay between the two components is expected but this is not seen in the 
data. A more severe constraint is that the energy of the second component is too low (it would be expected to
peak in the GeV range). A more plausible scenario is synchrotron emission from electrons accelerated in internal shocks. This scenario could be 
possible given that the width of this peak is not constrained to be as narrow as a Planck function. 

Finally, we note that it may also be that neither of the components arises from the photosphere. This would of course open the possibility of
many different origins, like the emission from two optically thin regions. However, it also means that there are few constraints; as discussed above, 
tying one component to the photosphere provides strong limitations to the origins of the other component. We therefore refrain from such 
speculations.

\section{Summary and Conclusions}

We have analysed the time-resolved spectrum of GRB 080825C using both the GBM and LAT instruments on {\fermi}. We find evidence for a 
second peak at energies above the previously reported Band component, present at the level of $3.5\sigma$. The temporal behaviour of this 
peak is different from components previously seen in GRB spectra; the peak energy increases throughout the emission episode, from a few hundred 
keV to several MeV. If real, this is the first time such a feature has been detected. 
The two components likely arise in different regions of the outflow, and assuming one of the components arises in the photosphere sets strong
constraints on the possible origins of the other. The most promising appears to be that the high-energy peak is photospheric emission in a
Poynting flux dominated outflow, where the {magnetisation increases} with time.

\section*{Acknowledgements}

We would like to thank Damien B{\'e}gu{\'e} for valuable comments and discussions.
MA is an International Research Fellow of the Japan Society for the Promotion of Science.

The \textit{Fermi}-LAT Collaboration acknowledges generous ongoing support
from a number of agencies and institutes that have supported both the development 
and the operation of the LAT as well as scientific data analysis. These include the 
National Aeronautics and Space Administration and the Department of Energy in the 
United States, the Commissariat \`a l'Energie Atomique and the Centre National de la 
Recherche Scientifique / Institut National de Physique Nucl\'eaire et de Physique des 
Particules in France, the Agenzia Spaziale Italiana and the Istituto Nazionale di Fisica 
Nucleare in Italy, the Ministry of Education, Culture, Sports, Science and Technology 
(MEXT), High Energy Accelerator Research Organization (KEK) and Japan Aerospace 
Exploration Agency (JAXA) in Japan, and the K.~A.~Wallenberg Foundation, the Swedish 
Research Council and the Swedish National Space Board in Sweden.
 
Additional support for science analysis during the operations phase is gratefully 
acknowledged  from the Istituto Nazionale di Astrofisica in Italy and the Centre National 
d'\'Etudes Spatiales in France.

\end{document}